# Enlarged scaling ranges for the KS-entropy and the information dimension

Holger Kantz
*Max-Planck-Institute for Physics of Complex Systems, Bayreuther Str. 40, D 01187 Dresden, Germany*

Thomas Schürmann
*Department of Theoretical Physics, University of Wuppertal, D 42097 Wuppertal, Germany*

1 August 1995

Numerical estimates of the Kolmogorov-Sinai entropy based on a finite amount of data decay towards zero in the relevant limits. Rewriting differences of block entropies as averages over decay rates, and ignoring all parts of the sample where these rates are uncomputable because of the lack of neighbours, yields improved entropy estimates. In the same way, the scaling range for estimates of the information dimension can be extended considerably. The improvement is demonstrated for experimental data.

## I. THE KS-ENTROPY

The unpredictability of a sequence of observations, let it be a time series representing a chaotic system or an encoded message, can be characterized by the Kolmogorov-Sinai (KS) entropy. It is the average additional information obtained by observing the state of the system at a certain instant, provided one has observed the whole past.[1] This amount of information per time step trivially is zero for a periodic sequence. Since uncorrelated random noises can not be predicted, the KS-entropy for such data equals the information transport rate through the observation channel, i.e. the logarithm of the number of possible different states (e.g., ln 2 for bit strings). Nontrivial values can be found for the output of systems with memory like Markov processes, but also for deterministically chaotic signals. Numerically, one computes entropies of finite order $m$, $h_m$ which in the limit of $m \to \infty$ converge to the KS-entropy. Unfortunately, as soon as $m$ becomes large with respect to the length of the time series, the estimates of $h_m$ are systematically underestimated and decay towards zero for even larger $m$, such that in many realistic chaotic systems $h_\infty$ cannot be estimated with reasonable precision. Moreover, certain definitions of the complexity of a sequence[2,3] rely on the way the $h_m$ converge towards $h_\infty$. Ebeling[3] postulates for certain complex systems a power law $h_m = h_\infty + \alpha m^{-\beta}$ with $\beta \approx 0.2-0.5$. For a numerical determination of $h_\infty$ and $\beta$ extremely large $m$ are required.

In this part the paper we report on a slight modification of the standard algorithm which yields enlarged scaling ranges in entropy estimates from a finite amount of input data.

We assume the data to be a scalar time series of real numbers $x_j$, $j = 1,\ldots,T$, representing a trajectory of a chaotic system. For the treatment of symbol sequences see below.

Assume a partition $\mathscr{P}$ of the state space. Denote by $p_{i_1,i_2,\ldots,i_m}$ the joint probability that a point of the trajectory lies in the $i_1$th element of the partition, its successor in time in the $i_2$th box, and so on. Then the block entropy of order $m$ reads

$$H_m = \sum -p_{i_1,\ldots,i_m} \ln p_{i_1,\ldots,i_m} \quad (1)$$

and the Kolmogorov-Sinai entropy

$$h = \sup_{\mathscr{P}} \lim_{m \to \infty} h_m = \sup_{\mathscr{P}} \lim_{m \to \infty} H_{m+1} - H_m. \quad (2)$$

The realization of the supremum over all partitions generally is nontrivial. In favourable cases generating partitions are known[4-6] and the string of real numbers can be converted into a symbol sequence. One can determine the probabilities in Eq. (1) by box-counting. Even in this favorable situation, in the limit of large block lengths $m$ one systematically underestimates the entropy. The reason and how to avoid this will become clear below.

In most situations, however, and in particular when dealing with experimental data, generating partitions are not known. One generally approximates the supremum by covering the attractor by a grid of boxes with edge length $\epsilon$, which becomes generating in the limit $\epsilon \to 0$. Unfortunately this yields an exploding number of joint probabilities in the limit $m \to \infty$ and Eq. (1) cannot be exploited numerically with reasonable effort.

One standard approach[7] consists in introducing an importance sampling originally employed by Grassberger and Procaccia[8] to determine the fractal dimension of attractors and later to find a lower bound of the KS-entropy, the correlation entropy.[9] Cohen and Procaccia[10] extended this method to the computation of the KS-entropy. Let $\vec{x}_i^{(m)} = (x_i, x_{i+1},\ldots,x_{i+m-1})$ be an element of the space of sequences of $m$ successive values of the trajectory. Count the number $n_i^{(m)}$ of $\epsilon$-neighbors of $M$ reference points $\vec{x}_i^{(m)}$ chosen randomly according to the invariant measure of the process. The estimate of the block entropy of block length $m$ is given by

$$H_m(\epsilon) = \ln(T-m+1) - \frac{1}{M} \sum_{i=1}^{M} \ln n_i^{(m)}(\epsilon), \quad (3)$$

and the estimate of the KS-entropy reads

H. Kantz and T. Schürmann: Enlarged scaling ranges

$$h_m(\epsilon) = H_{m+1}(\epsilon) - H_m(\epsilon)$$
$$= \frac{1}{M} \sum_{i=1}^{M} (\ln n_i^{(m)}(\epsilon) - \ln n_i^{(m+1)}(\epsilon)). \quad (4)$$

One has to perform the limit $m \to \infty$ and $\epsilon \to 0$ in order to find the asymptotic value of the KS-entropy. The inclusion of finite sample corrections obtained in Ref. 11 essentially amounts to substituting $\ln n$ by $\Psi(n+1)$, where $\Psi$ is the derivative of the gamma function. In particular, this allows for the possibility of $n$ to be zero. Alternatively, it is sometimes suggested to omit points without neighbours in Eq. (3), but this leads to wrong estimates of the block entropies.

By discussing the problem of reference points without neighbours we are already at the crucial point: The finite sample corrections regularize Eq. (3), but cannot supply the lacking information about the neighbour statistics. Therefore, in the limit that all reference points lose their neighbors Eq. (3) converges towards a constant and the estimate of $h$ becomes zero.

The last line of Eq. (4) shows the problem in a different light: As soon as a given reference point has no $\epsilon$-neighbor for some $m$ (i.e., $n_i^{(m)} = 0$), it will not have one for $m+1$, and the two terms cancel each other. Thus such a point does not contribute to the sum, but it is included in the normalization $1/M$. Therefore, the right hand side of Eq. (4) yields a systematically too small value for $h_m(\epsilon)$ for large $m$ and/or small $\epsilon$, which are the relevant limits. Obviously, we have the same problem when substituting the logarithm by the $\Psi$-function. Although our numerical results presented below always are based on the $\Psi$-functions, for convenience of notation we suppress in the following the finite sample corrections and deal with the logarithms. In all of the following expressions one can simply replace $\ln n$ by $\Psi(n+1)$ to incorporate them.

In Fig. 1 this well known problem of underestimation is demonstrated for a time series of a chaotic map (for details see sec. IV) with a KS-entropy of about 0.27 per time step. This decay towards zero extinguishes the scaling range [i.e., a regime where $h_m(\epsilon) \approx$ const] completely for those $\epsilon$ which are small enough to expect the true KS-entropy.

## II. ACCESSIBILITY SAMPLING

Equation (4) can be written as an expectation value:

$$h_m(\epsilon) = \langle \ln n_i^{(m)}(\epsilon) - \ln n_i^{(m+1)}(\epsilon) \rangle_M, \quad (5)$$

where $\ln n_i^{(m)} - \ln n_i^{(m+1)}$ is the logarithm of a decay rate. Thus the KS-entropy is the average of the decay (or escape) rate of neighbours when increasing the time window from $m$ to $m+1$.

As the main reason for the underestimation we isolated the problem that the difference $\ln n_i^{(m)}(\epsilon) - \ln n_i^{(m+1)}(\epsilon)$ is trivially bounded by $\ln n_i^{(m)}(\epsilon)$ (requiring $n_i^{(m+1)} \geq 1$). Thus the lack of neighbours does not predominantly introduce statistical fluctuations which could partially cancel each other, but a systematic underestimation, as soon as $\ln n_i^{(m)} < h$ [or $\Psi(n_i^{(m)}(\epsilon) + 1) < h$], i.e. as soon as for a given point less than about $e^h$ neighbours are found.

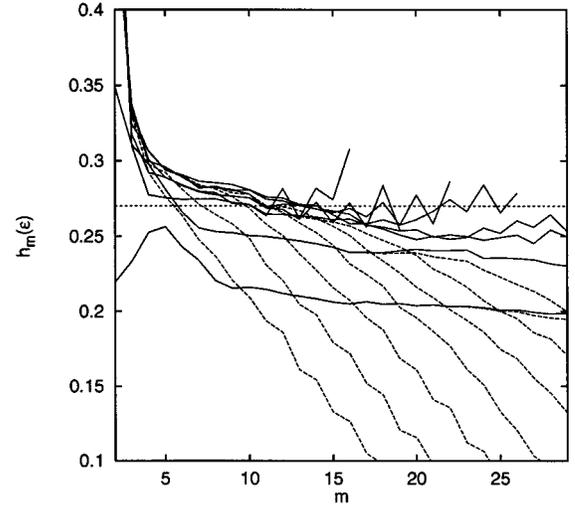

FIG. 1. Entropy estimates obtained by the Cohen-Procaccia correlation method Eq. (4) (broken lines) and the new reduced sampling Eq. (6) (continuous lines, truncated at the value of $m$ where the number of ignored reference points exceeds $\frac{2}{3}M$). The curves correspond to different values of $\epsilon$, ranging from $2^{-2}$ to $2^{-9}$ (from bottom to top). A trajectory of length $T = 100\,000$ of a chaotic map with $M = 2500$ reference points was used (for details about the map see sec. IV). The dotted line indicates the (only) positive Lyapunov exponent, $\lambda_+ \approx 0.27$.

The remedy we suggest is as simple as effective: Ignore all points without sufficiently many neighbours and compute

$$\widetilde{h}_m^{n_{\min}}(\epsilon) = \langle \ln n_i^{(m)}(\epsilon) - \ln n_i^{(m+1)}(\epsilon) \rangle_{n_i^{(m)} > n_{\min}} \quad (6)$$

for reasonable $n_{\min}$. Thus the importance sampling is substituted by an accessibility sampling. The improvement is demonstrated in Fig. 1 for the above used time series. Notice that this reduced sampling is different from just discarding contributions from Eq. (3).

Our modification can be criticized since it systematically changes the sample on which the average of the decay rates is obtained, in the way that all "thin" parts are ignored. In the following we want to show that this error is small, predominantly of statistical nature and can be controlled empirically.

It is obvious that in the limit of large $T$, $\widetilde{h}_m^{n_{\min}}(\epsilon)$ in Eq. (6) converges pointwise towards the true asymptotic value and in this limit is unbiased. However, it might be possible that for every finite $T$ there are still some systematic deviations [of course much smaller than in Eq. (3)] from the true value, since one ignores the decay rates $n_i^{(m)}/n_i^{(m+1)}$ on all parts of the space of $m$-vectors which are sampled insufficiently. The difference $\ln n_i^{(m)} - \ln n_i^{(m+1)}$ is the logarithm of the estimate of the local decay rate, whereas $n_i^{(m)}$ itself, which determines which reference points are excluded from Eq. (6), is $T$ times the estimate of the invariant measure in the space of $m$-vectors. A systematic error can only occur, if these two quantities are correlated. In fact they might be, since the lower the rate by which a reference point loses its neighbours when increasing $m$, the more likely this point remains in Eq. (6) for large $m$. Thus we cannot exclude that our way of sampling still contains some underestimation of the entropy when too many reference points are discarded. However, this error could be easily detected numerically: If the entropy is smaller than one bit per iteration, the average decay rate is smaller than one half. Thus in order to observe this decay with some confidence one needs not more than, say, $n_{\min}=4$ neighbours. Increasing $n_{\min}$ beyond 4 will not change the reliability of the estimated decay rates dramatically, but will further reduce the sample of reference points on which the average of Eq. (6) is based. Thus finding that $\widetilde{h}_m^{n_{\min}}(\epsilon)$ does not depend systematically on $n_{\min}$ as soon as $n_{\min} \geq 4$ (as we indeed do) indicates that the effect of ignoring the ''thin'' parts of the attractor is negligible [of course we observe statistical fluctuations if the number of remaining points in Eq. (6) becomes too small].

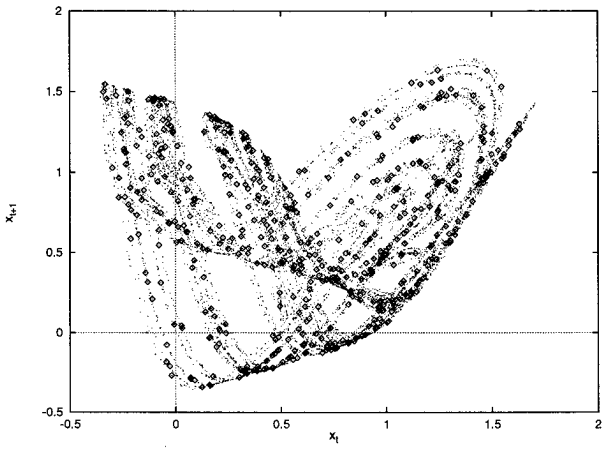

FIG. 2. Symbols: 500 reference points $\vec{x}_i^{(2)}$ on the Ikeda attractor, for which $n_i^{(11)} < n_{\min}$, i.e. which are excluded from the average for $h_{11}$ due to the accessibility sampling ($n_{\min}=3$, $T=20000$). The dots represent the attractor.

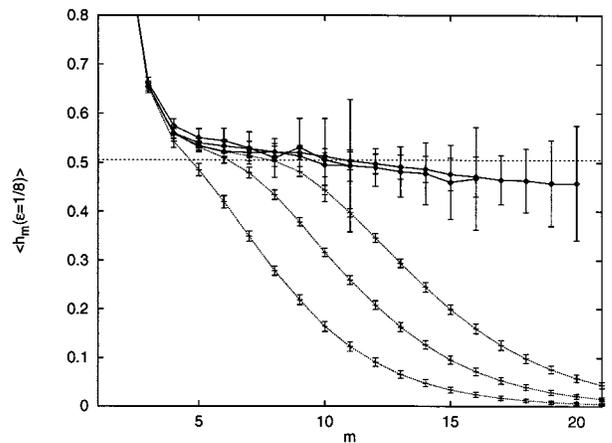

FIG. 3. Expectation values for a sample of 500 trajectories each of $\widetilde{h}_m^{n_{\min}=4}(\epsilon=0.125,T)$ (continuous lines) and $h_m(\epsilon=0.125,T)$ (broken lines) for trajectories of lengths $T=2048$, $T=8192$, and $T=32768$ (broken lines: from left to right, continuous lines: ending at $m=11$, $m=16$, $m=20$). The continuous curves are truncated, when the statistical error exceeds a reasonable bound. For comparison, we indicate the positive Lyapunov exponent $\lambda_+ = 0.505 \pm 0.002$ by the horizontal line. The data are the real parts of iterates of the Ikeda map $z_{t+1} = 1 + 0.9 z_t \exp(0.4i - [6i/(1+|z_t|^2)])$, $z_t \in \mathbb{C}$.

In Fig. 2 we present another test: On the attractor, the reference points which were discarded by our way of sampling are indicated. They turn out to be scattered unsystematically, thus rejecting the suspicion that one might sample the attractor systematically wrong.

Finally, we present in Fig. 3 the strongest argument in favour of our modification: We compute the averages of $\widetilde{h}_m^{n_{\min}=4}$ ($\epsilon=0.125$) over a huge ensemble of finite trajectories. The expectation value of $\widetilde{h}_m$ converges reasonably well towards the accurate value, thus demonstrating that if there is a bias it is small enough to not depreciate our method. The remaining underestimation of the entropy can, at least partially, be traced back to the fact that the value $\epsilon=1/8$ is still quite large and the induced partition thus still not generating. In principle, one therefore should compare the continuous curves to $h_m$ ($\epsilon=1/8$) computed for infinitely many data. Unfortunately, as one can see from Fig. 3, one should use a trajectory of at least $T \simeq 2^{25}$, which is beyond our possibilities. The same goes for Fig. 1. In conclusion, Fig. 3 demonstrates impressively that one can proceed to values of $m$ (and correspondingly of $\epsilon$) for which one would need orders of magnitude more data without our method.

## III. THE INFORMATION DIMENSION

One algorithm to compute the information dimension $D_1$ is again a generalization of the Grassberger-Procaccia algorithm[8] first published in Refs. 12 and 13. Computing again the number of neighbors $n_i^{(m)}(\epsilon)$ in an $m$-dimensional $\epsilon$-neighborhood of $M$ reference points, the information dimension is given by

$$D_1(\epsilon,m) \ln \epsilon \propto \frac{1}{M} \sum_{i=1}^{M} (\ln n_i^{(m)}(\epsilon) - \ln(T-m+1)). \quad (7)$$

If the data represent a fractal measure, one expects to find a range of $\epsilon$ and a value $m_0$, such that for $m > m_0$, $D_1(\epsilon,m) = d_1 =$ const. On the large scales generally no scaling can be found, since it is destroyed by the cutoff introduced by the finite extension of the attractor. Like the KS-entropy, $D_1$ decays to zero [again one should substitute $\ln n$ by $\Psi(n+1)$ to regularize the expression for vanishing $n$] in the limit of small $\epsilon$, such that for short data sets the scaling range may disappear. Rewriting Eq. (7) as

$$D_1 = \frac{1}{M(\ln \epsilon - \ln \epsilon')} \left( \sum_{i=1}^{M} \ln n_i^{(m)}(\epsilon) - \sum_{i=1}^{M} \ln n_i^{(m)}(\epsilon') \right)$$

$$= \frac{1}{M(\ln \epsilon - \ln \epsilon')} \sum_{i=1}^{M} (\ln n_i^{(m)}(\epsilon) - \ln n_i^{(m)}(\epsilon')) \quad (8)$$

reveals the same problem as discussed in connection with the KS-entropy: The relevant quantity determining the information dimension is the rate by which the number of neighbours decays when reducing $\epsilon$. Again, if the number of neighbors found for some $\epsilon$ is already too small, this decay rate cannot be computed any more and converges towards zero for points without neighbors. Thus we write Eq. (8) as an expectation value and base the average only on points with sufficiently many neighbors. Sufficiently many means $n_{\min} > (\epsilon/\epsilon')^{d_1}$, where of course $d_1$ is not known *a priori*, but may be estimated from the correlation dimension $d_2$. Thus we claim that

$$D_1(\epsilon,m) = \frac{1}{\ln \epsilon - \ln \epsilon'} \langle \ln n_i^{(m)}(\epsilon) - \ln n_i^{(m)}(\epsilon') \rangle_{n_i > n_{\min}} \quad (9)$$

yields by far enlarged scaling ranges for $D_1$.

The fixed mass method[13,14] offers a powerful alternative to the plain correlation method described by Eq. (8). However, by the introduction of the accessibility sampling, the correlation method becomes superior, since it allows one to proceed towards smaller length scales (see Fig. 4).

## IV. EXPERIMENTAL DATA

The improvement of entropy and dimension estimates becomes relevant when the data base is limited as in the analysis of experimental time series. Both quantities are of particular interest, since they allow for cross-checks, if one or more Lyapunov exponents of the system are also computed (see, e.g., Ref. 15). The information dimension should coincide with the Kaplan-Yorke dimension, $D_{KY} = k + (\Sigma_{i=1}^{k} \lambda_i / |\lambda_{k+1}|)$, where $k$ is the largest integer such that $\Sigma_{i=1}^{k} \lambda_i > 0$. The KS-entropy should equal the sum of the positive Lyapunov exponents (Pesin's identity). Substituting the KS-entropy and the information dimension by the correlation entropy and dimension, for whose computation very stable algorithms exist,[8] weakens the above relations to inequalities.

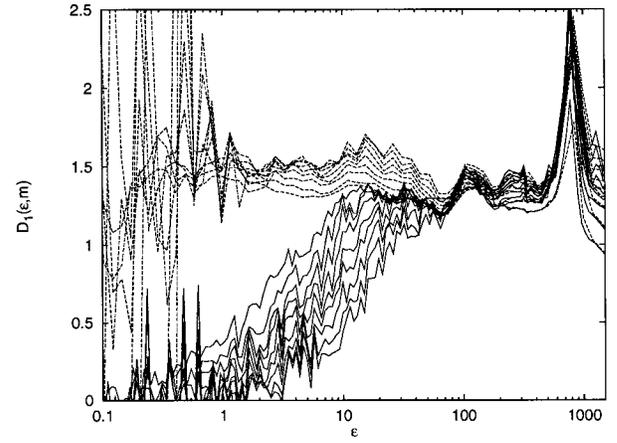

FIG. 4. The information dimension of the NMR laser data: continuous lines represent results obtained by Eq. (7), broken lines the ''accessibility sampling'' Eq. (9) ($m=2,...,10$). Note that for ideal low-dimensional data one would expect $D_1(\epsilon,m)$ to be independent of $m$ in the scaling range. The fact that these experimental data violate this may be due to a small amount of red noise.

Thanks to the help of the experimental group around Professor Brun at the ETH Zürich we possess a very fine time series of the output of an NMR laser.[16] In a proper Poincaré section these are 38000 points on an attractor of a dimension around 1.3 and a maximal Lyapunov exponent of about 0.3. After nonlinear noise reduction[17] the data are the best one can hope to find under realistic conditions to treat them by nonlinear diagnostics.

In Fig. 4 we present the estimates of the information dimension with and without our reduced sampling ($n_{\min}=2$).

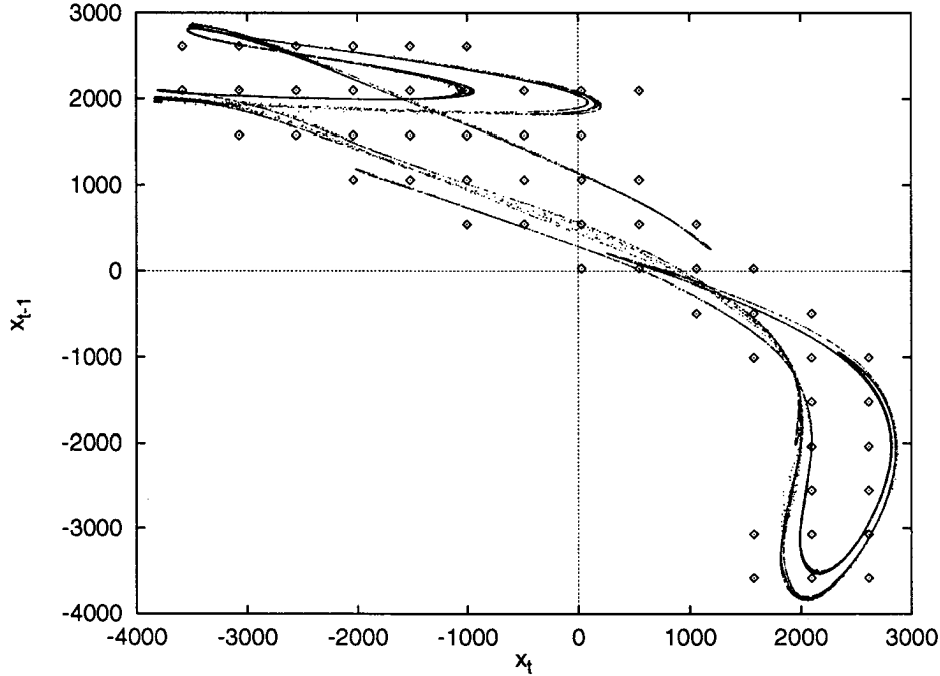

FIG. 5. The synthetic data obtained from the iteration of the fitted dynamics of the NMR laser data. The experimental data after noise reduction look almost indistinguishably the same. The symbols represent the (projections of) centres used for 140 radial basis functions of the Lorenzian type, $\phi(r) = 1/[1+(r/\alpha)^2]$ with $\alpha=1770$. With this fit the normalized one step prediction error $\langle (x_{t+1} - f(\vec{x}_t))^2 \rangle^{1/2} / \sigma_{\text{data}}$ is $4.5 \cdot 10^{-3}$.

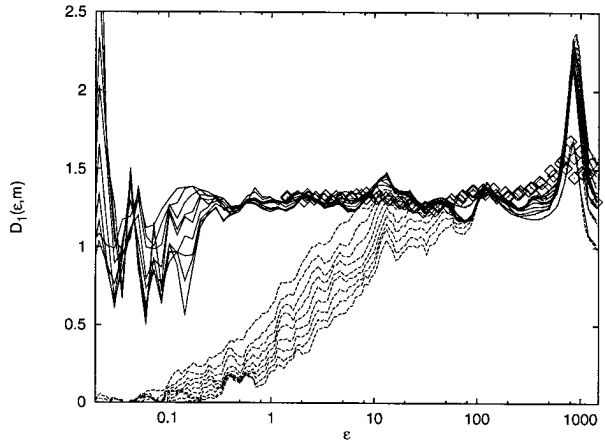

FIG. 6. Estimates of the information dimension as in Fig. 4, but for the noise-free synthetic time series of length 100000 gained by iterating the RBF fit $f$ ($m=2,...,10$). Symbols: results obtained by the fixed-mass algorithm for the same values of $m$, where $\langle \ln \epsilon_k \rangle = D_1 \ln(k/T)$ and $\epsilon_k$ are the distances between reference points and their $k$th neighbour in a time series of length $T$. In the plot, $\exp(\langle \epsilon_k \rangle)$ is plotted on the $\epsilon$-axis. Note that the smallest scale defined in this way is about one order of magnitude larger than the end of the scaling range for the accessibility sampling.

It is obvious that only with this large number of data we find an estimate already with Eq. (7), but that our way of sampling extends the scaling range by almost two orders of magnitude.

In order to clarify the problem of the slight $m$-dependence of $D_1$ and some $\epsilon$-dependence of $h$ (not shown) we create a synthetic noise-free data set. In a three-dimensional delay embedding space we fit the dynamics $f$ such that $x_{t+1}=f(\vec{x}_t)+r_t$, $\vec{x}_t=(x_t,x_{t-1},x_{t-2})$, by the method of radial basis functions (RBF).[18] Here, $r_t$ should be pure noise and minimal for a good fit of $f$. By minimizing the one-step-prediction error $e_0=\langle (x_{t+1}-f(\vec{x}_t))^2\rangle^{1/2}$ for the ansatz $f(\vec{x})=\Sigma_{j=1}^{140}\beta_j\Phi(|\vec{x}-\vec{y}_i|)$, where the radial basis functions are chosen as Lorentzians $\Phi(r)=1/[1+(r/\alpha)^2]$ and $\vec{y}_i$ are properly chosen fixed centres, one finds the coefficients $\beta_j$. Iterating $f$, starting with an initial condition in the vicinity of the observed data, yields an attractor which is extremely close to the one represented by the experimental data (Fig. 5). Thus we can generate a time series of this self-made dynamical system of arbitrary length, which should be characterized by the same invariants as the experimental data (a method called *bootstrapping* in Ref. 19), and at the same time we can compute the Lyapunov exponents of the model system by the standard method.[20] Repeating our analysis from above for a series of $T=10^5$ scalar values we gain Fig. 1 and Fig. 6, where in fact we find a clear scaling range for both $h$ and $D_1$. Thus we can summarize our numerical analysis by $D_1=1.3-1.35$, $h\approx 0.27$, and $\lambda_1\approx 0.27$, $\lambda_2\approx -0.699$, $\lambda_3\approx -1.2$, which yields $D_{KY}=1.38\pm 0.01$ in reasonable agreement with $D_1$.

## V. CONCLUSIONS

Ignoring points without sufficiently many neighbours in estimates of the KS-entropy and the information dimension avoids a prominent source for a systematic underestimation, and in all tested examples (logistic equation, Ikeda map, Hénon map, Lorenz system) yields promising results. This way of sampling requires one to two orders of magnitude less data than the conventional methods, thus allowing for estimates also in experimental time series. We presented arguments and numerical evidence that the fact that averages are not carried out over the correct sample does not introduce a too large bias. This way of reduced sampling can be easily employed also in box-counting algorithms, which for entropy estimates become important if one knows a binary generating partition. In this case one has to ignore sequences $i_1,i_2,...,i_m$ which occur not often enough.

## ACKNOWLEDGMENT

This work was supported by DFG within the Graduiertenkolleg ''Feldtheoretische und numerische Methoden in der Elementarteilchen und Statistischen Physik''.